\documentclass{ifacconf}

\usepackage{graphicx}      
\usepackage{natbib}        
\usepackage{amssymb}
\usepackage{amsmath}
\usepackage{graphicx}
\usepackage{subfigure}
\usepackage{caption}
\usepackage{psfrag}
\usepackage{balance}
\begin{document}
\begin{frontmatter}

\title{Toward Human-in-the-Loop Supervisory Control for Cyber-Physical Networks} 

\author{Mehdi Firouznia,} 
\author{Chen Peng,} 
\author{and Qing Hui}

\address{Laboratory for Engineered Network Systems (LENS), Department of Electrical and Computer Engineering, 
	University of Nebraska-Lincoln,  NE 68588-0511 (e-mail: mfirouznia2@ unl.edu, chen.peng@huskers.unl.edu, qing.hui@unl.edu).}

\begin{abstract}                
This work proposes a novel approach to include a model of making decision in human brain into the control loop. Employing the methodology developed in mathematical neuroscience, we construct a model that accounts for quality of human decision in supervisory tasks. We specifically focus on adaptive gain theory and the strategy selection problem. The proposed model is shown to be capable of explaining the change of a strategy from compensatory to heuristics in different conditions. We also propose a method to incorporate the effect of internal and external parameters such as stress level and emergencies in the decision model. The model is employed in a supervisory controller that dispatches the jobs between autonomy and a human supervisor in an efficient way.
\end{abstract}


\end{frontmatter}

\section{Introduction}

Supervisory control of complex cyber-physical networks with numerous parameters is a complicated and challenging task. Automated subsystems control the process in local and global levels, but human operator's overall supervision still plays a significant role. Human operator affects the system in different levels as a function of indigenous system states or exogenous factors such as emergencies.  Generally, it is assumed that in normal situation each plant can complete the designated task autonomously. However the trained human operator has the authority to take over the automation and issue commands that change plant state trajectories. This scheme requires the operator be constantly aware of the states of each plant and notice any anomaly sufficiently soon. Thus the issued corrections guarantee the safe operation. The condition that no longer holds during emergencies and the basic assumption that vindicates human operator's full authority are jeopardized. One approach to face this problem is to assign weights to autonomy and human decision based on the factors like threat levels (\cite{anderson2010optimal}). However it is hard to justify that higher level of autonomy in high threat levels would result in better control outcome or the other way around. Also since there is no human decision model involved, it is assumed that the trained operator in a perfect healthy condition would issue eligible and plausible commands.

The second approach is more involved. First it assumes that the system possess \textit{sliding levels of autonomy} so that the autonomy and human intervention interact dynamically. Second it employs the concept of \textit{human-in-the-loop} to design a complementary control, namely \textit{decision unit} which adds a model of human decision to the controller. The outcome of the decision unit can either be presented to the operator as a directive or may adjust the issued command toward better results. Thus far some research efforts have been conducted in the latter area using the model-predictive control (MPC) framework (\cite{MagnusMPC}). However, the most common approach toward constructing a tangible human decision model is to assume a given probability distribution for decisions in different conditions, based on previous recorded data (\cite{MagnusMPC,operartorchoice}).

In this research we will try to construct a model to capture the dynamics of making decision in human brain with the help of the methods developed in mathematical neuroscience. The problem of making decision in human brain has been widely studied in neuroscience community and a handful number of models both in the form of mathematical abstractions or biophysically connectionist models are available. In Section II we will briefly introduce them. The main contribution of this work is to include a cognitive model in the control loop in a way that it can be used to design a ``human-intelligent'' supervisory control system. Also, we will propose a method to incorporate the effect of work environment and human condition, such as stress or fatigue, into the proposed control framework.  

After reviewing the mathematical abstract methods that model simple two alternative choice tasks, we will propose our model for multi-choice multi-attribute tasks in Section II. In Section III we will use the framework of \textit{adaptive gain theory} to investigate the neuronal mechanisms that govern the process of making decision. We will explain the role of specific type of neurotransmitters that affects the quality of decision in brain. We will also define the process that changes the strategy of making decision specially in the high stress condition. In Section IV we will integrate the proposed model in the control loop. Specifically we will try to keep the human decision success rate to its optimum value by dispatching the task between anatomy and the human operator. 
Finally, Section V concludes the paper.

\section{Mathematical models of decision making}
\subsection{Single-cue two-choice task}
Mathematical abstract models to describe the process of making decision in human brain start by analyzing the  simple two-alternative forced choice (2AFC) decision task. Using the optimal sequential probability ratio test (SPRT), the process on its continuum limit converges to the famous drift-diffusion model (DDM) if symmetrical threshold is assumed. In that sense, the decision variable for a noisy evidence can be modeled by one-dimensional Wiener process bounded by positive and negative thresholds, $\theta_A$ and $\theta_B$, in which an integrator accumulates the difference of evidences between two choices (\cite{bogacz2006physics}). In order to reflect the effect of bounded accuracy and forgotten information, the DDM integrators are considered not perfect but leaky as follows: 
\begin{equation}
dx(t)=(\mu -\lambda x )dt + \sigma dW, \quad x(0)=0
\label{ddm}
\end{equation} in which $x(t)$ is the decision variable, $\mu$ is the drift, $\sigma$ is the diffusion rate, and $dW$ is the standard Wiener process. The term $\lambda$ represents the leak and $\lambda > 0$ leads to a stable Ornstein-Uhlenbeck (O-H) process. The drift rate $\mu$ represents the capability of input information in discrimination between choices A and B, which in behavioral studies are known as quality of cue (attribute). For example, in the random dot motion task designed by \cite{baker1982basis}, the cue is the direction of dots and its quality alters by changing the proportion of dots moving in the left or right direction. $\lambda$ tunes the drift based on the current state and is often connected to memory processes (e.g., primacy and recency effects), conflict situations (e.g., approach-avoidance), or similarities between choice alternatives (\cite{10.3389/fnhum.2014.00697}). In the free response paradigm, whenever the decision variable reaches the positive or negative thresholds, the decision is made, while in interrogation protocol, after spending the specific amount of time, the decision is made based on the value of gathered information. DDM has the capability to capture the characteristics of 2AFC in terms of speed or accuracy of decision.

\subsection{Multi-cue two-choice task}

For 2AFC, only one cue, e.g., the direction of dots in \cite{baker1982basis}, is concerned. In real world, always several cues, such as color, sound, and texture, are involved. One method is to combine and integrate all cues in favor of each choice into single source of evidence and this source is being used throughout the decision process. More involved treatment includes separate processes for each cue. In this approach the order of considering the cues and the process time devoted to each cue are two important aspects. The time frame of the decision process is divided to subintervals with different lengths during which the attention focus is only one cue. The order of cues can be assumed deterministic or probabilistic. Let $\mathcal{M} =\{1,2,\dots,M\}$ be the index set of cues.  Following the method of \cite{10.3389/fnhum.2014.00697} we assume that the evidence for cues is accumulated by a piecewise O-H process.

\begin{eqnarray}
dx(t) = \Big(\mu_m - \lambda_m x(t)\Big)dt + \sigma dW_m\\ \nonumber
m\in \mathcal{M}, \quad t_{l-1}\leq t< t_l
\label{cueOH} 
\end{eqnarray} in which the parameters identified by index $m$ are the characteristics of the $m$th cue.  If finite decision time span is divided to $L$ consecutive time intervals, [$t_{l-1},t_l$] for $l=1,\dots,L$, we assume that one cue is processed in each time interval based on the given order schedule. The process of making decision in free response time or interrogation is the same as a single cue task.    

\subsection{Single-cue multi-choice task}

In order to model multi-choice tasks, more general race model, which is comprised of separate leaky competing integrators, representing each choice, with mutual inhibition, was proposed by \cite{usher2001time}. Each integrator gathers information in favor or against the associated choice based on the value of cue. We assume that the dynamic of each integrator is governed by the O-H process.  Consider $K$  integrator pools, one for each choice, accumulating the incoming noisy evidence $S_i$ in favor of choice $i$. Each pool is described by the following form:
\begin{equation}
dx_i = \left(-kx_i-\sum _{j\neq i} w x_j +S_i\right)dt +\sigma_{i} dW_i
\label{racemodel}
\end{equation}
where $w$ is the mutual inhibition strength among pools. There are two different categories of decision criteria in choosing among the pools which give the same asymptotically optimal results (\cite{draglia1999multihypothesis}). One way is to assign a threshold $\theta_i$ to each pool and select the one that reaches the threshold sooner under the free response protocol or at the fixed time choose the pool with higher value of decision variable. Other methods consider the ratio of the output of the pools such as max-vs-next and max-vs-average (\cite{mcmillen2006dynamics}). \cite{bogacz2006physics} showed that under a particular parameter range, two-dimensional race model can be expressed as one dimensional DDM.

\subsection{Multi-cue multi-choice task}
In this paper, we propose the following leaky integrator race model to describe the dynamics of multi-cue multi-choice tasks. This model combines the race model and time and order scheduling concept as follows:

\begin{eqnarray}
dx_{i,m}(t)&=&\left(-k_mx_{i,m}(t)-\sum _{j\neq i} w_mx_{j,m} (t)+S_{i,m}\right)dt \nonumber\\
&&+\sigma_{i} dW_{i,m}, \quad m\in \mathcal{M}, \quad t_{l-1}\leq t< t_l
\label{multicurace}
\end{eqnarray} where task with index $m$ is selected according to cue ordering schedule and processed during the assigned time interval. The criteria of absolute threshold or max-vs- next or max-vs-average are considered to select the choice among the members of a decision vector stacked by $x_{i,m}$.

\subsection{Cue order schedule; Strategy selection}

The order of representing cues to the pools and the time interval during which the pools stay focused on the selected cue can be assumed to be deterministic or probabilistic based on given data. In the probabilistic version, uniform or normal distribution has been studied by \cite{10.3389/fnhum.2014.00697} without giving justification about the method of choosing the parameters. The question that we try to answer is what process controls the importance of a cue and hence the order and time span that a cue is presented to pools.  

Behavioral scientists have introduced normative models to explain the process of making decision in multi-cue tasks. In normative models each cue is assumed to have a utility or validity $[q_m]$ which is defined as the conditional probability that a choice based on this cue is correct, given that the cue discriminates between the choice alternatives (\cite{martignon2002fast}). Each cue also has a value $[c_{m,i}]$ in favor or against the choice represented by pool $i$ which for simplicity is assumed to be +1 or -1. Validity of each cue is multiplied by its value and the sum for each choice is calculated. The choice with the largest sum is selected. This approach is called weighted additive (WADD) and is defined as a subsection of larger category named compensatory methods (\cite{StrategyUnderpinning}) which in essence are the discretized versions of race model without noise, where cue validity is equivalent to binary drift rate. Another approach called non-compensatory or heuristics employs only a  limited number of cues and ignores others. Compensatory methods require extensive cognitive efforts and best results are expected when there is no time or information constraints. On contrary, when information is not enough or very noisy or the decision time is restricted, people are more pron to find a limited numbers of cues and decide according them, ignoring the value of other cues. The decision to select which of these two approaches is called   \textit{strategy selection}. In other words strategy is defined as a decision on how to make decisions.   

Returning to our multi-cue multi-task race model, the strategy selection problem is a process that controls the probability distribution of considering each cue and the time assigned to it in the order schedule. We employ the idea presented in  \cite{StrategyUnderpinning} to find such probability distribution. Assume that a weight factor is initially assigned to each cue on the basis of given initial cue validities. Then the weight factors are calculated according to the following softmax rule:
\begin{equation}
a_m =  \frac{e^{\gamma_E q_m}}{\sum_{i=1}^M e^{\gamma_E q_i}}
\label{weight}
\end{equation} in which $\gamma_E$ is the linearized excitatory gain factor. In the next section we will define $\gamma_E$ based on the neuronal activation function. We use the normalize vector $[a_m]$ as the probability distribution of cue processing time and order.
\begin{equation}
p_m = \frac{a_m}{\text{max}\  [a_m]}
\end{equation}  

Note that by increasing $\gamma_E$ the probability of choosing the cue with highest validity and the time interval assigned to it increases. As hypothesized in \cite{StrategyUnderpinning}, the gain can be interpreted as the control parameter that changes the strategy from compensatory to heuristics. 

\subsection{Biophysically connectionist models} 

Following the notion of \cite{bogacz2006physics}, the accumulation of evidence in the 2AFC task in lateral intraparietal cortex (LIP) is modeled by two competing, mutual inhibitory neuronal populations, each of them is selective to one alternative and described by: 
\begin{equation}
du_j=\left( -\lambda u_j-\sum_{k \notin i} \alpha f_{g,\gamma}(u_k) +I_j\right)dt +\sigma_{j} dW_j
\label{LIPmodel}
\end{equation}
where $u_{k}$ is the mean input current to each cell of the $i$th population, $\alpha$ is the mutual inhibition constant among populations, and $f_{g,\gamma}$ is the neuron activation function, mostly considered as a sigmoidal activation, $1/(1+\exp(-4\gamma (x-g)))$ with maximum slope $\gamma$ at point $g$. To make (\ref{LIPmodel}) mathematically tractable, the linearized version around maximum gain point of $f(\cdot)$ is considered as follows:
\begin{equation}
du_j=\left(-\lambda u_j-\sum_{k \notin i} \gamma_I u_k + I_j \right)dt +\sigma_{j} dW_j 
\label{linLIP}
\end{equation} in which $\gamma_I= \gamma \alpha$ is linear mutual inhibition gain and $I_{j}$ is the source of synaptic current drive coming from sensory mechanisms and polluted with noise. The dynamics of $I_{j}$ is defined by presynaptic processes that themselves include the activation function of neuronal pools in sensory part and recurrent activation of choice pools, which in the linearized version are related to $\gamma_E$, the linearized gain of excitatory connections.

Employing the dynamics of single neurons and inhibitory and excitatory synapses, \cite{wong2006recurrent} connected the abstract notion of leaky competing integrators to interconnected networks of four populations of neurons in the 2AFC framework. In addition to two selective populations, they considered one non-selective and one inhibitory inter-neurons. Their model reduction technique showed that
the firing rate of population of neurons in the mean-field sense follows the same dynamics described by (\ref{linLIP}), i.e., single neuron activity can be replaced by mean firing rate of population. This was a major breakthrough to relate the connectionist biophysically driven model to mathematical abstraction, i.e., the selective populations follow the dynamics described by race model (\ref{racemodel}).  
Using the 2-dimensional model they studied the qualitative behavior of the network near bifurcation points for 2AFC.

According to (\ref{linLIP}) the response of the system to stimuli in different conditions is governed by the value of neuronal excitatory and  inhibitory,  and $(\gamma_E, \gamma_I)$ gains. By defining the performance criteria of reward rate (RR), the interval of neuronal excitatory and inhibitory gains that lead to the best performance were studied in \cite{eckhoff2011dimension}. The results showed that by mutually changing the excitatory and inhibitory gains, a phenomenon known as \textit{neuromodulation}, an area of high performance decision making, exists in the excitatory-inhibitory gain plain. In the next section, we will investigate the neuronal mechanisms that control neuromodulation and effect the quality of decision.

\section{ Adaptive Gain Control Theory}
Intuitively, human performance on most tasks is best with an intermediate level of arousal and is worse with too little or too much arousal or stress. This inverted U-shape relationship is confirmed by the classic Yerkes-Dodson curve. The underlying brain mechanism that controls phenomena such as arousal and motivation is provided by several brain stem neuromodulatory nuclei with wide distribution and ascending projection to the neocortex. These neurons play crucial roles in cognitive behavior by releasing neurotransmitters, such as dopamine (DA), serotonin, and norepinephrine (NE). Any disturbance in such basic and pervasive functions cause trouble in cognition, emotion, and behavior. In addition to their direct effect on post synaptic neurons in the form of excitation or inhibition, these neurotransmitters modulate the effect of other neurotransmitters such as glutamate and gamma amino butyric acid (GABA) by change of neuronal gain or activity function.  Our focus in this study is to model the role of locus coeruleus (LC) neurons which release NE, and its different modes of activities during the process of making decision. In particular, we focus on \textit{adaptive gain control theory} proposed by \cite{aston2005integrative} and try to formulate it in the control framework.

Experiments showed two distinguished modes of activity in LC neurons . In \textit{tonic} mode, an elevated baseline activity is recognized without any bursts, while in \textit{phasic} mode bursts of activities have been recorded during moderate baseline. Accurate decision process are usually accompanied by bursts in the phasic mode (\cite{usher1999role}). By increasing the level of baseline activity in LC, engagement in the specific task and consequently the performance decrease. This is the LC tonic mode which is associated with more destructibility and greater tendency to respond to not relevant stimuli.  These findings lead to hypothesis that postulates LC phasic activity can be modeled as an temporal attentional filter that facilitates behavioral responses in the task-related decision process. However the information-processing function that the tonic mode may serve needs more speculation. 
%

\begin{figure}
	\begin{center}
		\includegraphics[width=8cm]{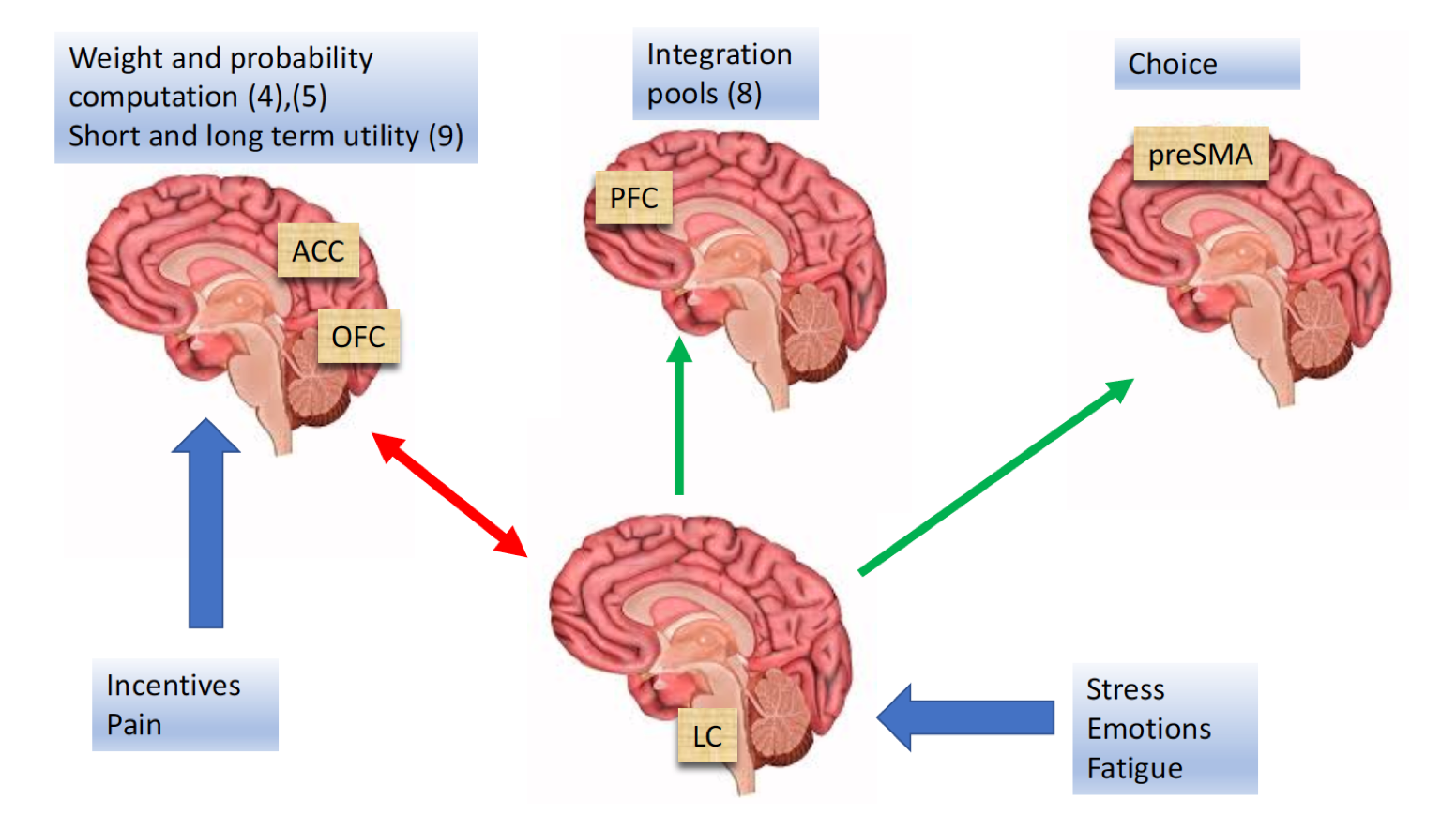}    
		\caption{Brain parts that govern decision process, the numbers in parenthesis indicate the equation number.} 
		\label{LC Controller}
	\end{center}
\end{figure}

\subsection{LC-NE Framework }
Studies on neuromodulatory effect of NE on the performance of making decision by \cite{eckhoff2011dimension} showed that the decision network can move from unaroused states through high performance to impulsive states and eventually lose the inhibition driven race model behavior by varying the gain of inhibitory and excitatory neurons. Furthermore, after making the decision in decision layer, the temporary increase in the gain $\gamma_E$ of neurons in behavioral layer makes them act as a binary units to reduce the effect of noise and delay in eliciting response. In other word, the phasic bursts make the multilayer complex decision network to act as a single layer network when there exists a strong evidence in favor of the decision. In contrast, LC tonic mode, due to elevated baseline activity, causes indiscriminate persistent increase in gain that renders more sensitivity to irrelevant stimuli. With respect to the current task, such distraction is clearly disadvantageous, however it paves the road for exploration of other opportunities and accumulate evidence toward other decisions.

Using a detailed population-level model of LC neurons and abstract connectionist network,  \cite{usher1999role} showed that change in electrical coupling among LC neurons can produce the abovementioned two modes of activity. Within LC, increased coupling gain, resulted from activation of the target decision unit, facilitates phasic mode, which in turn causes the alternation of gain function in neurons receiving NE in each layer of the behavioral network, namely input, decision layer itself, and response layer. This positive feedback loop leads to better performance. In contrast, reduced coupling strength in LC neurons causes a modest increase in baseline activity due to non-decisive random input into the LC neuron and diminished bursting activity. Employing this mechanism, decision maker optimizes the performance in a broader sense, which is the tradeoff between exploitation of current utility or exploration of other opportunities. The question is what will drive the gain change or baseline excitatory drive to change the mode of LC neurons? 

LC neurons have strong projection on two other frontal brain areas, namely orbitofrontal cortex (OFC) and anterior cingulate cortex (ACC), which are responsible for evaluation of reward and cost. OFC neurons are activated by rewarding stimuli and their response varies in proportion to the amount of reward. ACC is responsive to a variety of negatively valenced signals from pain to perceived errors in performance, in addition to task difficulty and conflicts in processing. The main feature which plays a significant rule in decision making is the ability of integration of rate of reward over extended periods.  Due to mutual projection between LC and OFC and ACC, it has been suggested that OFC and ACC may drive the LC activation in phasic mode. They also regulate the transition between phasic and tonic modes. When evaluations in ACC and OFC indicate that the current task utility decreases steadily, they facilitate transition to the tonic mode to search for other possible sources instead of focusing on the current task. This is done by diminishing the phasic bursts that render concentration on the current task. In case that ACC evaluates the utility is adequate enough, the phasic mode would continue. 

\begin{figure*}
	\centering

	\subfigure[Schematic Diagram]{\includegraphics[width=0.45\textwidth]{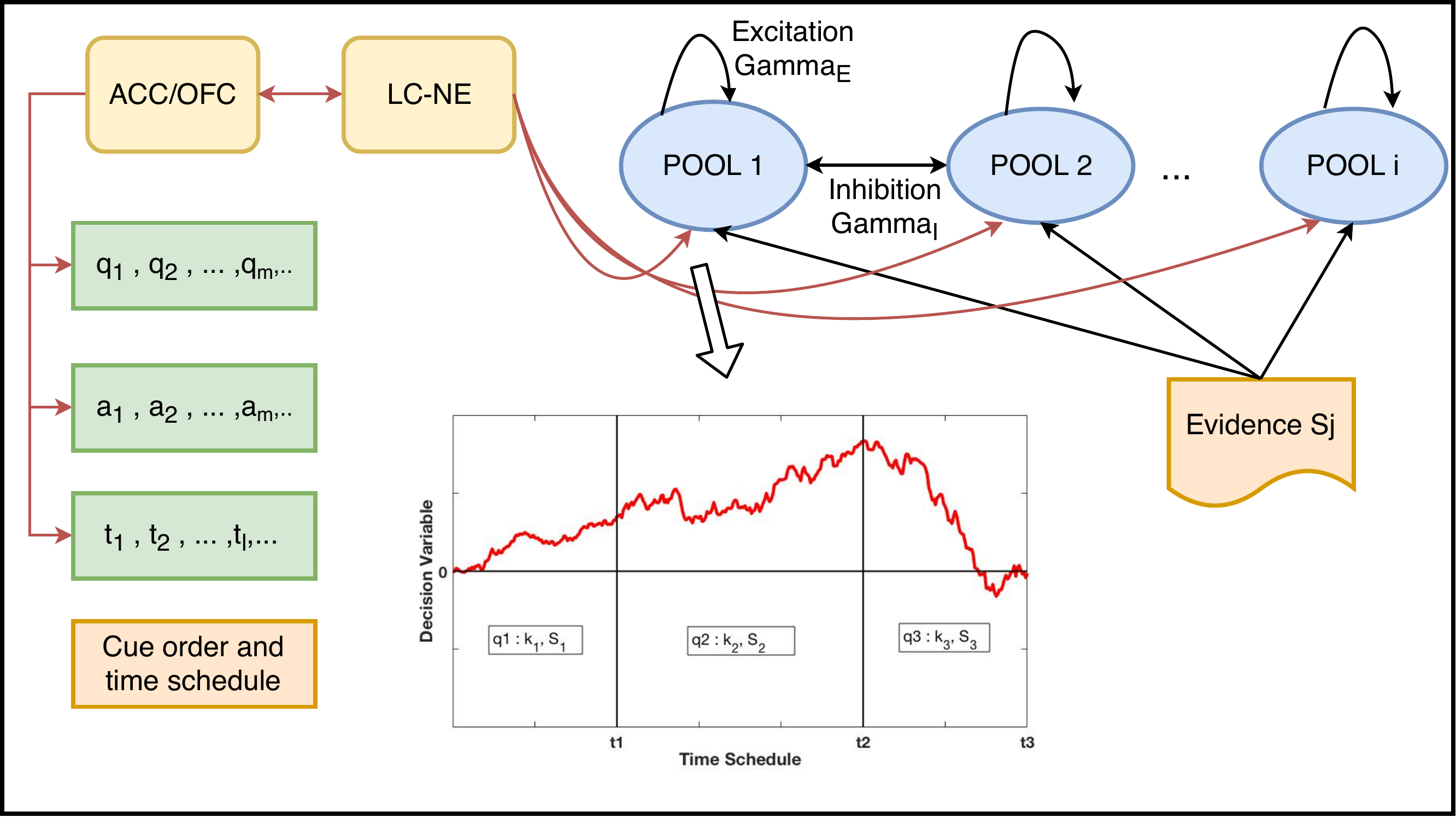}}~~~
	\subfigure[Block Diagram]{\includegraphics[width=0.50\textwidth]{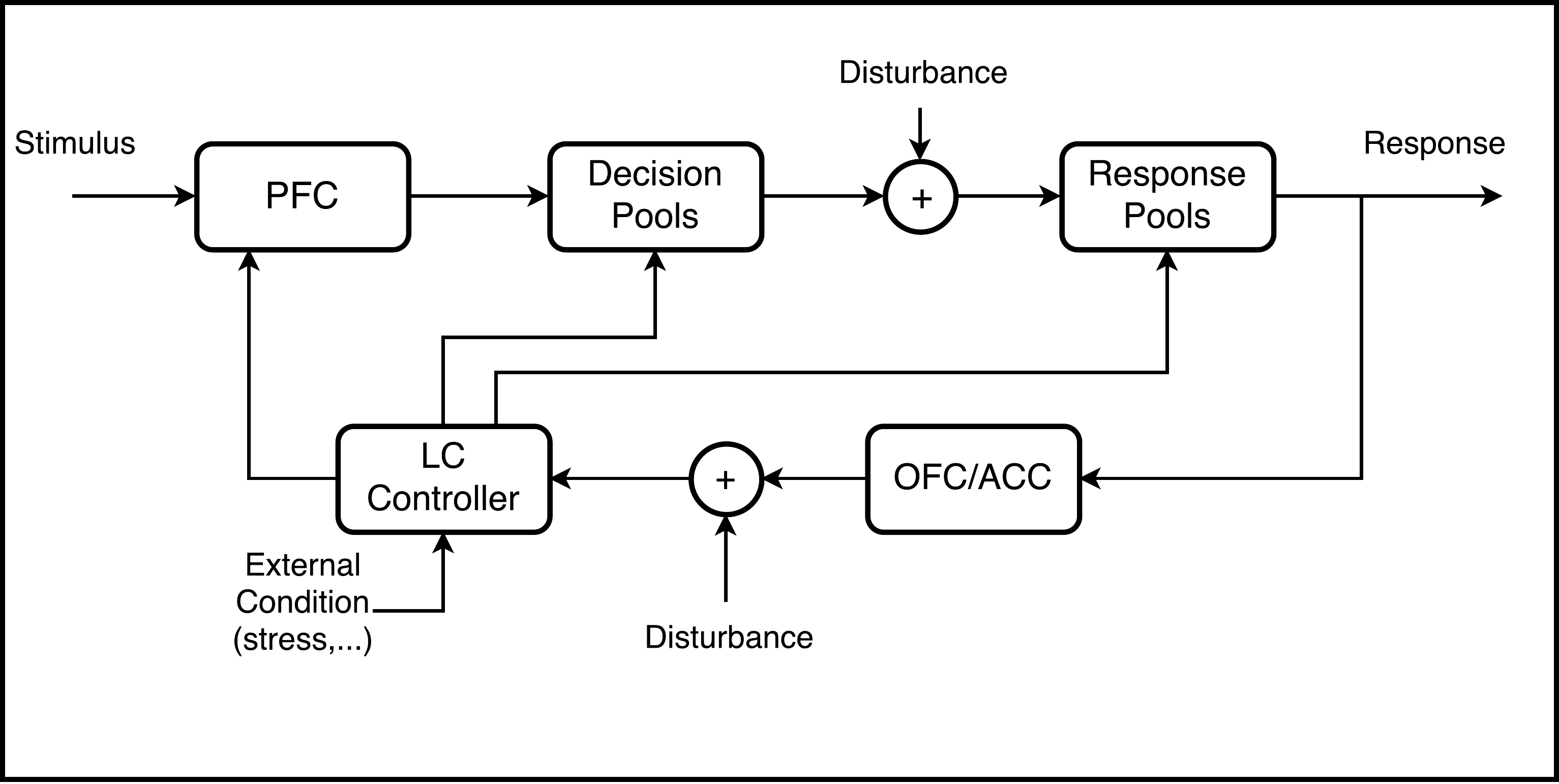}}
	\caption{Proposed adaptive gain control and strategy selection framework.} \label{fig:framework}
\end{figure*}
\begin{figure}
	\begin{center}
		\includegraphics[width=8cm]{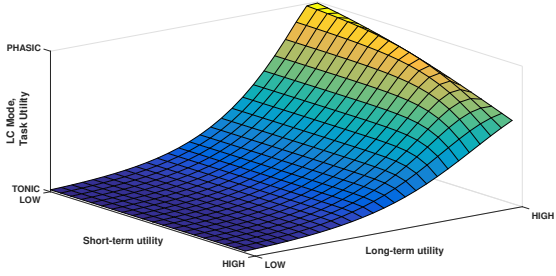}    
		\caption{Change of LC mode based on long and short term utilities.} 
		\label{utility}
	\end{center}
\end{figure}

The dilemma of either exploitation of current resource or exploration of other opportunities requires both long-term and short-term evaluation of utility. It is assumed that OFC and ACC outputs are integrated over two time scales in order of seconds and minutes respectively. For example for 2AFC a simple engagement index defined as:
\begin{equation}
E_i = \left(1-\frac{1}{1+\exp(u_s)}\right)\left(\frac{1}{1+\exp(u_l)}\right)
\end{equation} in which $u_l$ and $u_s$ are long and short term utilities of the decision. $E_i$ sets a threshold for changing the mode of LC neurons as depicted in Fig. \ref{utility}.

To fit the strategy selection into adaptive gain theory, it is hypothesized  that ACC and OFC also calculate the cue weight factor $[a_m]$ on the basis of initial cue validities $[q_m]$ (\cite{StrategyUnderpinning}). LC neurons on their phasic mode facilitate  adjustments of the weights factors in favor of most valid cues by tuning the linear gains $\gamma_E$ and $\gamma_I$. It has been believed that attentional controls such as cue order schedule and processing time are performed in prefrontal cortex (PFC) (\cite{rossi2009prefrontal}). Hence, the cue order schedule and process time distribution are determined in PFC. 
The projection of NE to all these areas ratifies the role of LC neurons in attention and making decisions. Figures \ref{LC Controller} and \ref{fig:framework} summarize the extended adaptive gain theory with strategy selection layer. 

\section{Human-in-the-Loop Supervised Controller}
In this section, we will employ the decision model to design a high level supervisory system that dispatches the jobs between anatomy and operator. This system is aware of both system and human conditions and its ultimate goal is to keep the success rate around its maximum value. 
First, we assume that a sequence of tasks $\mathbf{m}_1,\mathbf{m}_2,\dots$ arrive at the control center one after another, where $\mathbf{m}_i$ is a vector containing all relevant information of task $i$, and we do not discriminate between using $i$ or $\mathbf{m}_i$ when referring to task $i$. Furthermore, it is assumed that the complexity of task $\mathbf{m}_i$ can be estimated and its inverse simplicity and is given by a scalar $m_i$. Let us consider the case where only one human operator is in charge of completing all the tasks. In addition, each task can be completed by the human operator or done autonomously. The control center needs to decide whether a task should be assigned to the human or the autonomous computational model. The completion of each task has only two possible outcomes, i.e., success or failure. The success rate $p_1(\mathbf{m}_i)$ associated with the computational model for each task $\mathbf{m}_i$ is determinant and is assumed to be given. The success rate $p_0(\mathbf{m}_i,\Gamma_i)$ of the human operator not only depends on the specific task $\mathbf{m}_i$, but also his/her internal state represented by the value of $\Gamma$. As mentioned previously, we restrict ourselves to  the LC-NE system, in particular we represent operator internal states by the variable  $\Gamma_i=\{\gamma_I^i,\gamma_E^i\}$ at the time of task $\mathbf{m}_i$, 
where $\gamma_I^i$ is the inhibitory gain and $\gamma_E^i$ is the excitatory gain of neurons.
The human operator's success rate $p_0(\mathbf{m}_i,\Gamma_i)$ is estimated using the reward rate $R(\gamma_I^i,\gamma_E^i)$ given by :
\begin{equation}
\text{Reward Rate} = \frac{<\text{Acc}>}{\text{DT}+\text{NDT}+\text{RSI}}
\label{eq:RR}
\end{equation}
where $<$Acc$>$ is the fraction of trails with correct decision, DT is the time duration from the onset of stimulus to making the decision in brain. NDT is the sensory and motor delays and RSI is the time between trails.
Reward rate is scaled by the simplicity $m_i$ of task $\mathbf{m}_i$, i.e., $p_0(\mathbf{m}_i,\Gamma_i)=m_iR(\gamma_I^i,\gamma_E^i)$.

\begin{figure}[thpb]
	\centering
	\framebox{\parbox{3in}
		
		\includegraphics[scale=.3]{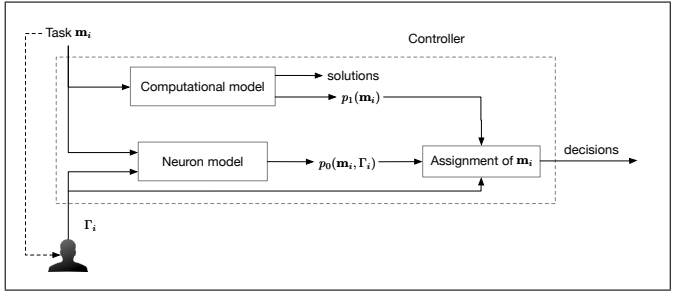}}
	\caption{Human-in-the-Loop Supervisory Controller}
	\label{fig:supervized}
\end{figure}

Next, we assume that the assignment of tasks affects the human operator's performance in the following way. If the human operator is assigned task $\mathbf{m}_i$, his/her internal state $\Gamma_i$ will diverge from the region of high performance by a random distance $\delta \Gamma_i^0$ after completing the task; if the human operator skips task $\mathbf{m}_i$, his/her internal state will be brought closer to the region by  a random $\delta \Gamma_i^1$. Finally, the controller design is given as follows. The probability $p$ of assigning task $\mathbf{m}_i$ to the computational model is given by 
\begin{equation}
p(\mathbf{m}_i,\Gamma_i)=(1-p_0(\mathbf{m}_i,\Gamma_i))p_1(\mathbf{m}_i)
\end{equation}
Thus the probability of assigning task $\mathbf{m}_i$ to the human operator is 
\begin{equation}
1-p=p_0(\mathbf{m}_i,\Gamma_i) + (1-p_0(\mathbf{m}_i,\Gamma_i))(1-p_1(\mathbf{m}_i))
\end{equation}
It can be seen that this controller prefers using human when $p_0$ is high, and will be more likely to use the computational model when the human has not enough success rate compared to the computational model. Fig.~\ref{fig:supervized} summarizes the proposed human-in-the-loop supervisory control framework.
\begin{figure*}[thpb]
	\psfrag{human success rate p0}[c][c][0.5]{human success rate $p_0$}
	\psfrag{average success rate p}[c][c][0.5]{average success rate $\bar{p}$}
	\psfrag{task i}[c][c][0.5]{task $i$}
	\centering
	~\\
	\subfigure[ $\gamma_E$ and  $\gamma_I$ gain trajectory]{\includegraphics[width=0.31\textwidth]{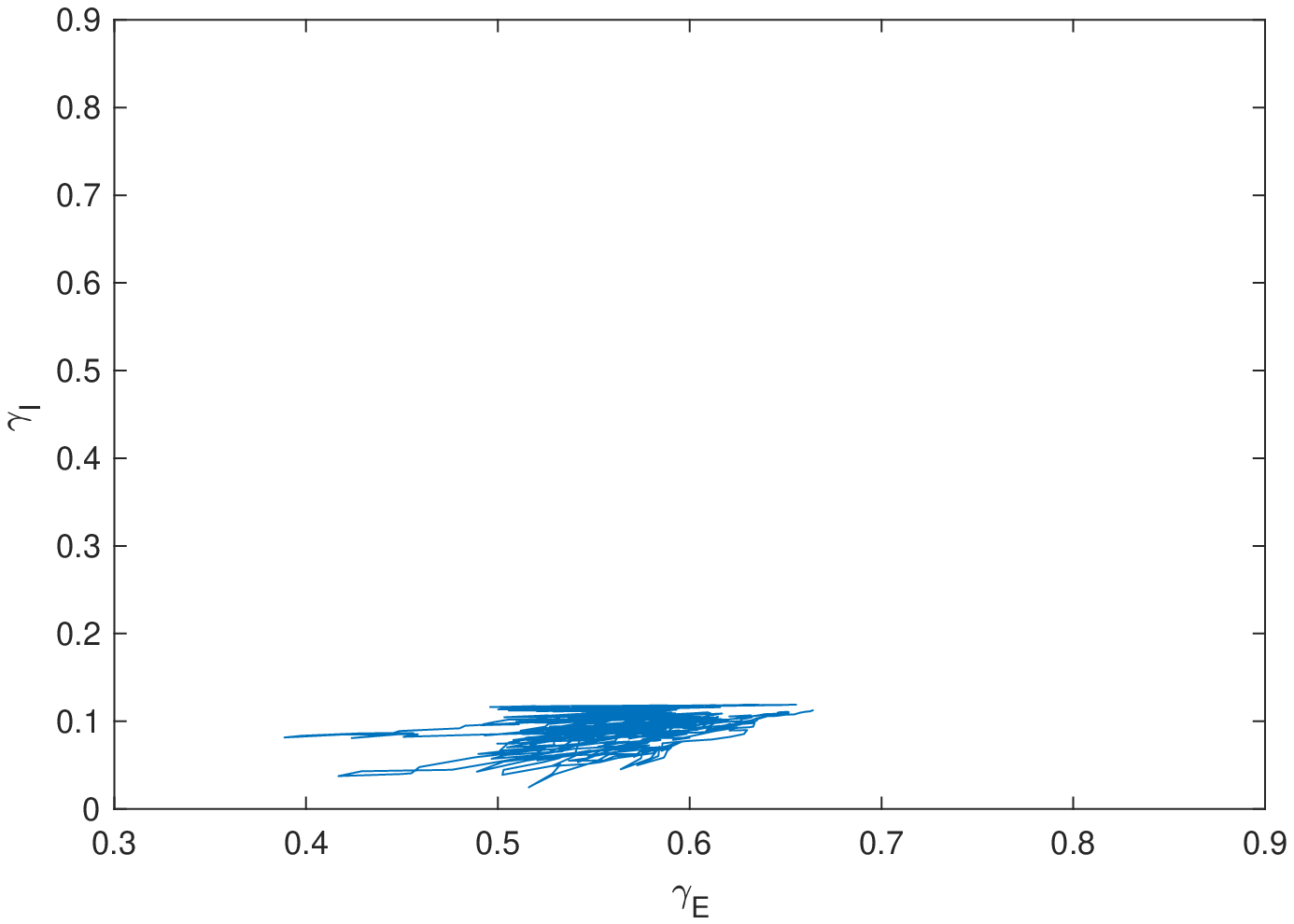}}
	\subfigure[Human operator's success rate $p_0$]{\includegraphics[width=0.31\textwidth]{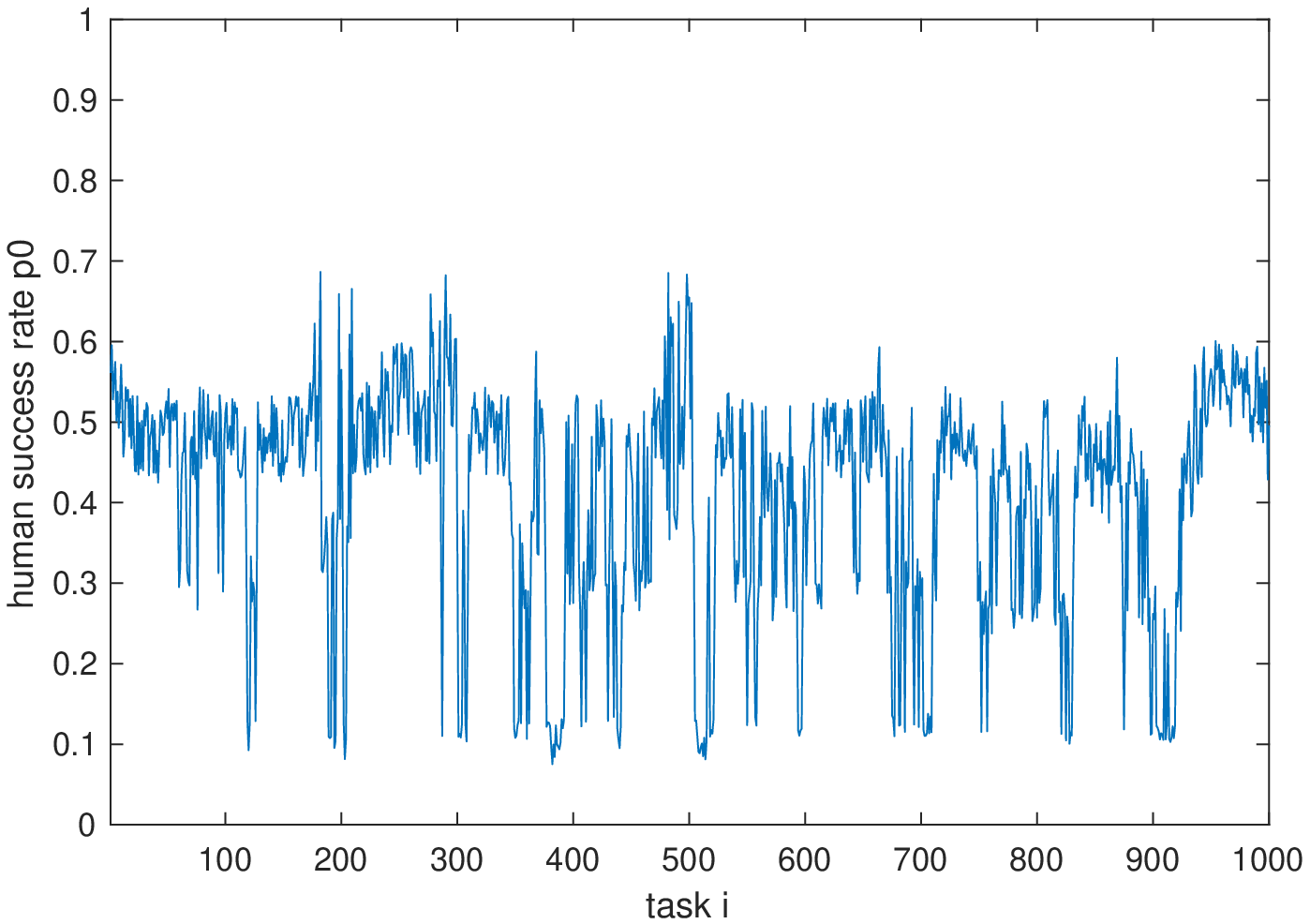}}
	\subfigure[Average success rate $\bar{p}$]{\includegraphics[width=0.31\textwidth]{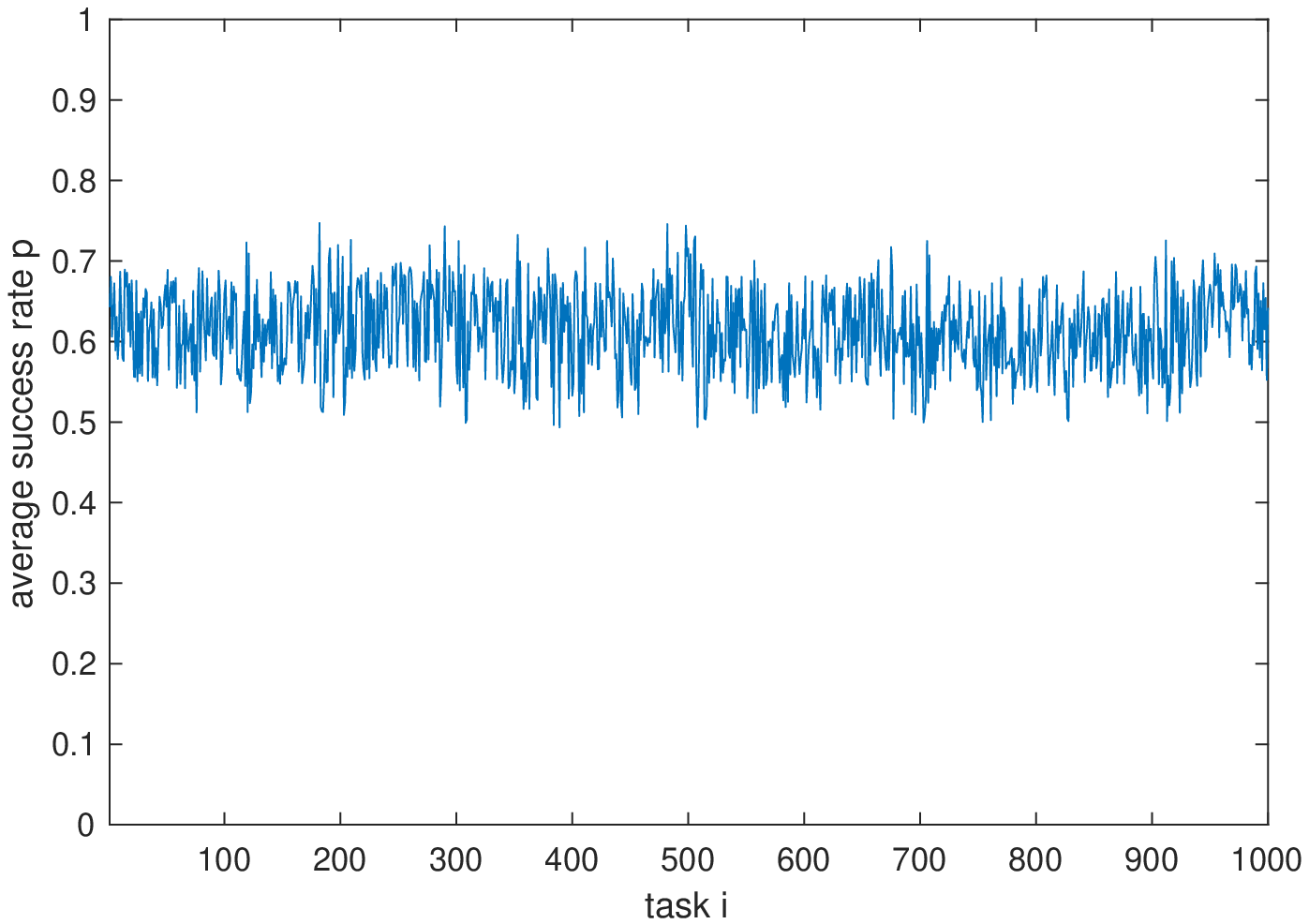}}
	\caption{Performance of human operator and average success rate under  supervisory control} \label{fig:exp_res}
\end{figure*} 

 \section{Simulation}
In this section we will simulate the effect of the proposed supervisory controller in the simple 2AFC task. The operator is required to make a decision based on the single cue, e.g., to detect a direction of motion correctly.  The  neuronal model consists of three types of neuronal population: primary visual cortex (V1) that gathers the information $S_i$, the middle temporal area (MT), and the lateral intraparietal cortex (LIP). The detail description of the model can be found in \cite{Beyeler2014}.  Visual stimulus is given in the form of gray scale video feed and the performance of  the model in detection of the direction  is analyzed. To obtain the best performance area in the $\Gamma$ plane, we simulate the reward rate of the  detection task over the range of inhibitory and excitatory gains. The aim of the controller is to keep the operator performance in the vicinity of high performance region.  We assume that when the operator is at best condition, his/her success rate is approximately $95\%$ for doing a simplest task, and $75\%$ for doing a most complex task. Since the highest reward rate $R(\gamma_I^i,\gamma_E^i)$ given by the simulated model is normalized to $1$, the task simplicities $m_i,i=1,2,\dots$ is uniformly randomly generated in the range $[0.75,0.95]$. 
Furthermore, we assume that the computational model success rate is given by $p_1(m_i)=0.95*m_i$, such that the success rate for the autonomous computational model doing a simplest task $m_i=1$ is $90\%$. 

The neural model simulation was ran on Tesla$^{\textregistered}$ K40 GPU with K40MGPUs on each node in the Holland Computing Center (HCC) at University of Nebraska-Lincoln (UNL). The MATLAB$^{\textregistered}$ 2017a script was run on iMAC$^{\textregistered}$, intel$^{\textregistered}$ Core$^{\mathrm{TM}}$ i5, with 8 GB memory and Mac OS X version 10.12.5.   
The experimental result is shown in Fig.~\ref{fig:exp_res}. It can be seen from Fig.~\ref{fig:exp_res}(a) that the system ensures that the $\gamma_E$ and  $\gamma_I$ gains are kept close to the region of best performance. As the operator continues to work, his/her success rate fluctuates, as shown in Fig.~\ref{fig:exp_res}(b), which is a representative situation under emergencies. The average success rate can be computed as 
\begin{equation*}
\bar{p}=(1-p)p_0 + p p_1
\end{equation*}
and is shown in Fig.~\ref{fig:exp_res}(c). It is clear that the average success rate is kept above 0.5, and with a much smaller variance due to the supervisory controller.

Note that the assumption that operator state (i.e.,  $\gamma_E$ and  $\gamma_I$ gains pair) diverges from its optimal position when the operator handles a task and converges when the operator skipped a task is naturally established, and is crucial to forming the control loop; however, the specific movement of the state is not important for our controller design. In fact, the controller needs only to know the consequences (divergence/convergence) of its actions (i.e., assign a task to the operator or computational model), and the current success rate estimation. The operator's success rate is thus where the controller is linked to the neuronal decision model. As compared in Fig.~\ref{fig:exp_res}(b) and (c), through proper arrangement of the assignment of tasks and a decent computational model, a higher and much stabler average success rate can be achieved. 
From here it can be speculated that with more computational models, each with a unique success rate profile, the average success rate can be lower.
However, a practical difficulty can be the estimation of computational model success rate.
Thus, in addition to searching for more accurate operator cognitive models,
future research directions on the controller can include studying the specification of complexity/simplicity of tasks, estimation of the computational model success rate, and theoretical aspects regarding the stochastic stability of such supervised controller.
\section{Conclusion}
\label{concl}
We proposed a prototype for a supervisory controller that includes the dynamics of making decision by humans in the control loop. A two-stage model for strategy selection and decision making was utilized to cover a range of situations, in which the operators are required to make proper decisions. The process of making decision in the brain was modeled with the well-known LC-NE framework.  A controller was designed to dispatch the tasks between the system and the operator to keep the operator close to the best performance area. A case study for a simple one-attribute task was simulated to show the effectiveness of the proposed controller. Future research would include more diverse normative models for selecting strategies and more complicated connectionist models that include a variety of biophysical factors.

\section*{Acknowledgment}
The authors would like to thank Dr. Adam Caprez in the HCC at UNL for his help and advice regarding neural model simulation. 

\balance

\bibliography{ifacconf}             

\begin{thebibliography}{17}
\providecommand{\natexlab}[1]{#1}
\providecommand{\url}[1]{\texttt{#1}}
\providecommand{\urlprefix}{URL }
\expandafter\ifx\csname urlstyle\endcsname\relax
  \providecommand{\doi}[1]{doi:\discretionary{}{}{}#1}\else
  \providecommand{\doi}{doi:\discretionary{}{}{}\begingroup
  \urlstyle{rm}\Url}\fi

\bibitem[{Anderson et~al.(2010)Anderson, Peters, Pilutti, and
  Iagnemma}]{anderson2010optimal}
Anderson, S.J., Peters, S.C., Pilutti, T.E., and Iagnemma, K. (2010).
\newblock An optimal-control-based framework for trajectory planning, threat
  assessment, and semi-autonomous control of passenger vehicles in hazard
  avoidance scenarios.
\newblock \emph{International Journal of Vehicle Autonomous Systems}, 8(2-4),
  190--216.

\bibitem[{Aston-Jones and Cohen(2005)}]{aston2005integrative}
Aston-Jones, G. and Cohen, J.D. (2005).
\newblock An integrative theory of locus coeruleus-norepinephrine function:
  adaptive gain and optimal performance.
\newblock \emph{Annu. Rev. Neurosci.}, 28, 403--450.

\bibitem[{Baker~Jr and Braddick(1982)}]{baker1982basis}
Baker~Jr, C.L. and Braddick, O.J. (1982).
\newblock The basis of area and dot number effects in random dot motion
  perception.
\newblock \emph{Vision Research}, 22(10), 1253--1259.

\bibitem[{Bertuccelli and Cummings(2012)}]{operartorchoice}
Bertuccelli, L.F. and Cummings, M.L. (2012).
\newblock Operator choice modeling for collaborative uav visual search tasks.
\newblock \emph{IEEE Transactions on Systems, Man, and Cybernetics - Part A:
  Systems and Humans}, 42(5), 1088--1099.
\newblock \doi{10.1109/TSMCA.2012.2189875}.

\bibitem[{Beyeler et~al.(2014)Beyeler, Richert, Dutt, and
  Krichmar}]{Beyeler2014}
Beyeler, M., Richert, M., Dutt, N.D., and Krichmar, J.L. (2014).
\newblock Efficient spiking neural network model of pattern motion selectivity
  in visual cortex.
\newblock \emph{Neuroinformatics}, 12(3), 435--454.
\newblock \doi{10.1007/s12021-014-9220-y}.

\bibitem[{Bogacz et~al.(2006)Bogacz, Brown, Moehlis, Holmes, and
  Cohen}]{bogacz2006physics}
Bogacz, R., Brown, E., Moehlis, J., Holmes, P., and Cohen, J.D. (2006).
\newblock The physics of optimal decision making: a formal analysis of models
  of performance in two-alternative forced-choice tasks.
\newblock \emph{Psychological review}, 113(4), 700.

\bibitem[{Chipalkatty et~al.(2013)Chipalkatty, Droge, and
  Egerstedt}]{MagnusMPC}
Chipalkatty, R., Droge, G., and Egerstedt, M.B. (2013).
\newblock Less is more: Mixed-initiative model-predictive control with human
  inputs.
\newblock \emph{IEEE Transactions on Robotics}, 29(3), 695--703.
\newblock \doi{10.1109/TRO.2013.2248551}.

\bibitem[{Diederich and Oswald(2014)}]{10.3389/fnhum.2014.00697}
Diederich, A. and Oswald, P. (2014).
\newblock Sequential sampling model for multiattribute choice alternatives with
  random attention time and processing order.
\newblock \emph{Frontiers in Human Neuroscience}, 8, 697.
\newblock \doi{10.3389/fnhum.2014.00697}.

\bibitem[{Draglia et~al.(1999)Draglia, Tartakovsky, and
  Veeravalli}]{draglia1999multihypothesis}
Draglia, V., Tartakovsky, A.G., and Veeravalli, V.V. (1999).
\newblock Multihypothesis sequential probability ratio tests. i. asymptotic
  optimality.
\newblock \emph{IEEE Transactions on Information Theory}, 45(7), 2448--2461.

\bibitem[{Eckhoff et~al.(2011)Eckhoff, Wong-Lin, and
  Holmes}]{eckhoff2011dimension}
Eckhoff, P., Wong-Lin, K., and Holmes, P. (2011).
\newblock Dimension reduction and dynamics of a spiking neural network model
  for decision making under neuromodulation.
\newblock \emph{SIAM journal on applied dynamical systems}, 10(1), 148--188.

\bibitem[{Martignon and Hoffrage(2002)}]{martignon2002fast}
Martignon, L. and Hoffrage, U. (2002).
\newblock Fast, frugal, and fit: Simple heuristics for paired comparison.
\newblock \emph{Theory and Decision}, 52(1), 29--71.

\bibitem[{McMillen and Holmes(2006)}]{mcmillen2006dynamics}
McMillen, T. and Holmes, P. (2006).
\newblock The dynamics of choice among multiple alternatives.
\newblock \emph{Journal of Mathematical Psychology}, 50(1), 30--57.

\bibitem[{Rossi et~al.(2009)Rossi, Pessoa, Desimone, and
  Ungerleider}]{rossi2009prefrontal}
Rossi, A.F., Pessoa, L., Desimone, R., and Ungerleider, L.G. (2009).
\newblock The prefrontal cortex and the executive control of attention.
\newblock \emph{Experimental brain research}, 192(3), 489.

\bibitem[{Usher et~al.(1999)Usher, Cohen, Servan-Schreiber, Rajkowski, and
  Aston-Jones}]{usher1999role}
Usher, M., Cohen, J.D., Servan-Schreiber, D., Rajkowski, J., and Aston-Jones,
  G. (1999).
\newblock The role of locus coeruleus in the regulation of cognitive
  performance.
\newblock \emph{Science}, 283(5401), 549--554.

\bibitem[{Usher and McClelland(2001)}]{usher2001time}
Usher, M. and McClelland, J.L. (2001).
\newblock The time course of perceptual choice: the leaky, competing
  accumulator model.
\newblock \emph{Psychological review}, 108(3), 550.

\bibitem[{Wichary and Smolen(2016)}]{StrategyUnderpinning}
Wichary, S. and Smolen, T. (2016).
\newblock Neural underpinnings of decision strategy selection: A review and a
  theoretical model.
\newblock \emph{Frontiers in Neuroscience}, 10, 500.
\newblock \doi{10.3389/fnins.2016.00500}.

\bibitem[{Wong and Wang(2006)}]{wong2006recurrent}
Wong, K.F. and Wang, X.J. (2006).
\newblock A recurrent network mechanism of time integration in perceptual
  decisions.
\newblock \emph{Journal of Neuroscience}, 26(4), 1314--1328.

\end{thebibliography}
                                                   







\end{document}